\documentstyle[12pt]{article}
\font\titolo=cmbx12 scaled\magstep2

\font\tsnorm=cmr12

\font\tscorsp=cmti10
\def\CQG{{\em Class. Quant. Grav. }}

\def\NPB{{\em Nucl. Phys. }}
\def\PLB{{\em Phys. Lett. }}  
\def\PRL{{\em Phys. Rev. Lett. }}
\def\PRD{{\em Phys. Rev. }} 
 
\def\MPLA{{\em Mod. Phys. Lett. }} 
 
\def\z{Z\kern -4.6pt Z}
\def\dx{\int d^3x\ \sqrt{-g}\ }
\def\l{\lambda}
\def\ha{{1\over 2}}
\def\a{\alpha}
\def\g{\gamma}
\def\d{\delta}
\def\f{\phi}
\def\ef{e^{-2\phi}}
\def\m{\mu}
\def\n{\nu}
\def\r{\rho}
\def\s{\sigma}
\def\x{\chi}
\def\ds{ds^2=}
\def\ku{k+1}
\def\kv{k+2}

\def\ra{\rightarrow}
\def\la{\l^2}
\def\i{\infty}
\def\be{\begin{equation}}
\def\ee{\end{equation}}
\def\bea{\begin{eqnarray}}
\def\eea{\end{eqnarray}}
\def\bc{\begin{displaymath}}
\def\ec{\end{displaymath}}
\def\lb{\label}
\begin{document}
\pagestyle{empty}
\null
\vskip 5truemm
\begin{flushright}
INFNCA-TH9610\\
june 1996
\end{flushright}
\vskip 15truemm
\begin{center}
{\titolo THE DUALITIES OF 3D DILATON GRAVITY}
\end{center}
\vskip 15truemm
\begin{center}
{\tsnorm Mariano Cadoni}
\end{center}
\begin{center}
{\tscorsp Dipartimento di Scienze Fisiche,  
Universit\`a  di Cagliari,}
\end{center}
\begin{center}
{\tscorsp Via Ospedale 72, I-09100 Cagliari, Italy.}
\end{center}
\begin{center}
{\tscorsp and}
\end{center}
\begin{center}
{\tscorsp  INFN, Sezione di Cagliari.}
\end{center}
\vskip 15truemm
\begin{abstract}
\noindent
We investigate Brans-Dicke dilaton gravity theories in 
2+1 dimensions. We show that the reduced field equations for 
solutions with diagonal metric and depending only on one spacetime
coordinate have a continuous $O(2)$ symmetry.
Using this symmetry we derive  general  
static and cosmological solutions of the theory.
The action of the discrete group $O(2,\z)$ on the space of the
solutions is discussed. Three-dimensional string 
effective theory and three-dimensional general relativity 
are discussed in detail. In particular, we find that the previously 
discovered 
black string solution is dual to a  spacetime with a conical
singularity.  
\end{abstract}
\vfill
\begin{flushleft}
{\tsnorm PACS:  04.60.kz, 04.70.-s, 11.25.-Pm\hfill}
\end{flushleft}
\smallskip
\vfill
\hrule
\begin{flushleft}
{E-Mail: CADONI@CA.INFN.IT\hfill}
\end{flushleft}
\vfill
\eject
\pagenumbering{arabic}
\pagestyle{plain}
\section{Introduction}
\paragraph{}
Target space duality is a very useful concept in string theory 
\cite{GPR}. It 
relates one with the other exact string backgrounds, which from 
the field 
theoretical point of view seem to be drastically different. 
For instance,
the duality symmetry interchanges singularities with horizons 
\cite{AG,VV} or, for 
cosmological string backgrounds \cite {GV,GAV,TV}, it relates 
expanding universes with contracting ones. 

Though duality symmetry is a genuine string feature that cannot be 
fully 
understood in terms of a purely (field theoretical) low-energy string 
theory, one can also analyze it in the context of a string effective 
theory. In this framework, the duality symmetries have a natural 
interpretation in terms of discrete symmetry transformations acting 
on the field 
theoretical degrees of freedom. This approach has been very useful 
because, on one hand, one  can establish connections between 
well-known field 
theoretical dualities and target space duality (for instance the  
electric-magnetic  duality of four dimensional vector fields). On the 
other hand, one can generalize the concept of  
duality symmetry to theories whose solutions  do not 
have a direct interpretation in terms of exact string backgrounds, 
i.e as  
two-dimensional  conformal field theories. For instance, it has been 
shown that the duality symmetries of the two-dimensional (2D) exact 
string 
black hole or cosmological backgrounds have a natural 
generalization in 
terms of the symmetries of black hole and cosmological solutions 
of 2D dilaton gravity  models \cite{CM, CC}. Moreover, 
it has been found 
that scale factor 
duality is a general feature of isotropic and  homogeneous 
cosmological 
solutions of  $d$-dimensional Brans-Dicke  dilaton 
gravity theories \cite{JL}. 

The purpose of  this paper is to analyze, from the point of view of 
duality symmetries,  three-dimensional (3D)  Brans-Dicke  
dilaton gravity theories. This class of models contains, as particular 
cases, the two most popular models of gravity in three dimensions:
3D  general relativity with a cosmological constant, whose black hole 
solutions were found by Ba\~nados, Teitelboim and Zanelli (BTZ) 
\cite{BTZ} and 3D string effective theory.

Three-dimensional dilaton gravity  has been already studied by 
several 
authors \cite{SKL, CHM, JLE}. Even though the black hole solutions 
of the theory 
have been already found, a general description of its static and 
cosmological 
solutions is still lacking.
In this paper we first show that for a metric with a diagonal 
form the 
requirement of independence of  
the solutions 
from two spacetime coordinates 
 implies the existence of a continuous $O(2)$
symmetry of the reduced field equations.
In particular, this is true for static, circularly  symmetric, 
solutions 
 and for
homogeneous (i.e. space-independent) cosmological solutions.
The $O(2)$ symmetry  implies the existence of a 
discrete $O(2,\z)$
symmetry group  acting on the space of the solutions,  which 
can be considered as a subgroup of the $O(2,2,\z)$ group. 
This discrete 
group appears in the context of string theory both as duality 
symmetry for 2D 
toroidal compactifications  and as symmetry of a curved, exact
3D string background independent of two 
spacetime coordinates (see for example Ref. \cite{GPR} and 
references 
therein). The existence of the continuous $O(2)$ symmetry is  
related to the solubility of the field equations. In fact, 
we will be able 
to find the most general static, circularly symmetric 
and homogeneous cosmological 
solutions of our 3D dilaton gravity models.
As a by-product we will also obtain the general solutions of 
this kind for
 3D string 
effective theory and we will find that a duality transformation 
maps the 
previously discovered  
black string \cite{HH} into a 3D spacetime with a conical singularity.  

The outline of our paper is as follows. The continuous $O(2)$ and the 
discrete $O(2,\z)$ symmetries of 3D dilaton gravity are derived and 
described 
in section 2. The static, circularly symmetric, solutions of the 
theory are 
derived and discussed in section 3,  where in particular the action of 
the duality transformations on the spectrum of the solutions and 3D 
string effective theory are  analyzed. In section 4 we consider 
cosmological solutions; the general solution together with some 
interesting particular cases  is derived and discussed. Finally,  
section 5 illustrates our conclusions. 
 
\section{The duality transformations}
\paragraph{}
Let us consider Brans-Dicke dilaton gravity in 2+1 dimensions. 
The action has the
following form:
\be 
S={1\over 2\pi}\dx\ef( R-4\omega(\nabla\f)^2+\l^2),
\lb{e1}
\ee
where $R$ is the scalar curvature, $\f$ is the dilaton, $\l$ is a 
constant and $\omega$ is the (3D) Brans-Dicke parameter.
We will trade  $\omega$ for a new parameter $k$,
\begin{displaymath}
\omega=\left({1\over k}-1\right), 
\end{displaymath}
to make our formulae simpler.
To each value of $k$ corresponds a different dilaton gravity theory.
For example, $k=\i$ corresponds to the 3D low-energy string effective 
action,
 for $k=0$ one gets 3D general relativity with a cosmological 
 constant.

The field equations stemming from the action (\ref{e1}), 
 for $k\neq-1$, can be written after some manipulation as follows:
\begin{displaymath}
 R_{\m\n}+2\nabla_\m\nabla_\n\f-{4\over k}\partial_\m \f\partial_\n 
 \f+{\l^2\over k+1}g_{\m\n}=0,
\end{displaymath}
\be \lb{e2}
\nabla^2\ef={k\over k+1} \la \ef.
\ee
For $k=-1$ consistency of the field equations requires $\l=0$. In the 
following we will take $k\neq -1$.  

In this paper, we will consider only those solutions of the field 
equations for which  the 3D spacetime 
metric has 
a diagonal form and that depend only on one spacetime coordinate. 
This is the case of static, circularly symmetric, solutions. 
Using the gauge freedom relative to 
transformations of the radial coordinate, we can  
write  the most general static, circularly symmetric, solution  
in the following form:
\be \lb{e3}
\ds-e^{2\rho(\s)}dt^2+d\s^2+e^{2\n(\s)}d\varphi^2,\quad 0\le\varphi
\le 2\pi,\qquad \f=\f(\s).
\ee
With this ansatz, the field equations (\ref{e2}) become
\bea \lb{e4}
\n''&+&\n'^2+\n'\r'-2\n'\f'-{\la\over k+1}=0,
\nonumber\\
\r''&+&\r'^2+\n'\r'-2\r'\f'-{\la\over k+1}=0,
\nonumber\\
\n''&+& \r''+\n'^2+\r'^2-2\f''+ {4\over k}\f'^2-{\la\over k+1} =0,
\nonumber\\
\f''&-&\f'\left(2\f'-\n'-\r'\right)=-\la {k\over 2(k+1)},
\eea
where $'=d/d\s$.
The field equations (\ref{e4}) are invariant under the action of a 
$O(2)$ group transforming the fields $\r,\n, \f$.
To show this assertion, we first note that the equations (\ref{e4}) 
are equivalent to the 
 equations of motion obtained from the action 
\be \lb{e5}
S=-\int d\s e^{-2\f+\r+\n}\left[4\r'\f'+4\n'\f'-2\n'\r'
+4\left({1\over k}-1\right)\f'^2-\l^2\right],
\ee
by varying $S$ with respect to $\r,\n$ and $\f$, together with the 
Hamiltonian constraint 
\be \lb{e6}
4\r'\f'+4\n'\f'-2\n'\r'
+4\left({1\over k}-1\right)\f'^2+\l^2=0.
\ee
The lagrangian in eq. (\ref{e5}) can be obtained  
by substituting the ansatz (\ref {e3}) into the action (\ref{e1}),
after dropping the total derivative terms, which do not 
contribute to the 
equations of motion.
For $k\neq -2$ let us define the new variables
\bea\lb{e7}
Z&=&\sqrt{{k+2}\over k} \tilde{Z}+{2\over \sqrt{k(k+2)}}\, U,
\nonumber\\
\tilde{Z}&=& \n+\r, \qquad Y=\n -\r,\nonumber\\
U&=& 2\f-\n-\r.
\eea
Expressed in terms of $Z,Y,U$, the action (\ref{e5}) and the constraint 
(\ref{e6}) become respectively
\be
S=\int d\s e^{-U}\left[ {k+1\over k+2}\, U'^2 -{1\over 2}
\left(Z'^2+Y'^2\right)+\l^2\right], \lb{e8}
\ee
\be 
{k+1\over k+2}\,U'^2 -{1\over 2}
\left(Z'^2+Y'^2\right)-\l^2=0.\lb{e8a}
\ee
It is now easy to see that the action (\ref{e8})  and the constraint 
(\ref{e8a}) are 
invariant under transformations of the group $O(2)$, which leave $U$ 
invariant and 
act on  the variables $Z$ and $Y$. This group contains, as subgroups, 
the rotations in the $(Z,Y)$ plane:
\begin{equation}
\left(\begin{array}{cc}Z\\Y\end{array}\right)\rightarrow
\left(\begin{array}{cc}
\cos\beta &-\sin \beta \\
\sin \beta & \cos\beta
\end{array}
\right) \left(\begin{array}{cc}Z\\Y\end{array}\right) \;,\lb{e9}
\end{equation}
with $0\le\beta\le 2\pi$, and the two inversions
\be\lb{f1}
\begin{array}{c}
Y\rightarrow -Y\\
Z\ra Z\end{array}\; ,\qquad 
\begin{array}{c}
Y\rightarrow Y\\
Z\ra -Z\end{array}\; .
\ee

We are particularly interested in the discrete $O(2,\z)$ subgroup 
of the $O(2)$ group. This discrete symmetry 
of the field equations (\ref{e4}) is clearly related to 
the $O(2,2,\z)$ 
symmetry of  
the exact 3D string backgrounds independent of two  spacetime 
coordinates (see Ref. \cite{GPR} and references therein).
  As we shall see later, when we  will discuss 3D string 
effective theory,
this $O(2,\z)$  symmetry  
group should be 
considered as the generalization of 3D target space string duality 
symmetry 
to the  generic dilaton gravity theory defined by the action 
(\ref{e1}).

The $O(2,\z)$ group is generated  by five elements, 
the two inversions (\ref{f1})
and  three rotations (\ref{e9}) with $\beta=\pi/2, \pi, 3\pi/2$.
Using eqs. (\ref{e7}), (\ref{e9}) and (\ref{f1}) one can easily 
obtain the representation 
of the $O(2,\z)$ group in terms of transformations of the 
variables $\r,\n,\f$.
We will give here only the representations for the two 
inversions and 
the $\beta=\pi$ rotation. These  transformations represent 
three independent subgroups of the $O(2,\z)$ group; each 
transformation 
$T$ satisfies $T^2=I$, where $I$ is the identity.  They can be 
therefore 
considered as duality transformations  of the field equations that, 
as we will see
in the next section, interchange different solutions of the theory.
The $\beta=\pi$ rotation and the two inversions (\ref{f1}) are 
represented, 
respectively, by the following transformations:
\bea\lb{f4}
&\r&\ra {1\over k+2}\left( 2\n-k\r -4\f\right),\nonumber\\
&\n&\ra{1\over k+2}\left( 2\r-k\n -4\f\right),\nonumber\\
&\f&\ra{1\over k+2}\left( -k \r-k\n +(k-2)\f\right);
\eea

\bea\lb{f3}
\r\ra\n,\nonumber\\
\n\ra\r,\nonumber\\
\f\ra\f;
\eea
\bea\lb{f2}
&\r&\ra{1\over k+2}\left( 2\r-k\n -4\f\right),\nonumber\\
&\n&\ra{1\over k+2}\left( 2\n-k\r -4\f\right),\nonumber\\
&\f&\ra{1\over k+2}\left( -k \r-k\n +(k-2)\f\right).
\eea

The previous discussion applies for every $k\neq -2$. It is also 
interesting to work out in detail the particular cases $k=0$ and 
$k=\i$.
For $k=0$ the action (\ref{e1}) reduces to 3D general relativity 
with a 
cosmological constant. Here the dilaton must be a constant, which 
 can be set equal to zero without loss of generality. The 
field equations are now invariant only under the inversion $Y\ra-Y$, 
the 
duality (\ref{f3}) with $\f=0$.
The $k=\i$ case corresponds to 3D low-energy string effective theory 
so that one would expect the dualities (\ref{f4}), (\ref{f3}), 
(\ref{f2}) 
to coincide with the
3D target space dualities of string theory.
For $k=\i$ the eqs. (\ref{e8}), (\ref{e8a}) become 
\bea
S&=&\int d\s e^{-U}\left[ U'^2 -{1\over 2}
\left(Z'^2+Y'^2\right)+\l^2\right], \nonumber\\
 U'^2 &-&{1\over 2}
\left(Z'^2+Y'^2\right)-\l^2=0,\nonumber
\eea
whereas the five elements of the $O(2,\z)$ group are represented, 
in terms of
transformations acting on the fields $\r,\n,\f,$ as follows
\be\lb{f6}
\begin{array}{c}\r\ra-\r\qquad \quad\\ \n\ra-\n\qquad\quad\\
\f\ra \f-\n-\r\end{array}\; ,\qquad  
\begin{array}{c}\r\ra\n\\ \n\ra\r\\
\f\ra \f\end{array}\; ,\qquad  
\begin{array}{c}\r\ra -\n\qquad\quad\\ \n\ra-\r\qquad\quad\\
\f\ra \f-\n-\r\end{array}\; ,
\ee
\be\lb{f7} 
\begin{array}{c}\r\ra -\n\quad\\ \n\ra\r\qquad\\
\f\ra \f-\n\end{array}\; ,\qquad  
\begin{array}{c}\r\ra\n\qquad\\ \n\ra-\r\quad\\
\f\ra \f-\r\end{array}\; .  
\ee
 The transformations (\ref{f6}) are the dualities of the field 
 equations 
 (\ref{e4}) in the $k=\i$ case and can be obtained by taking 
 this limit 
 in eqs. (\ref{f4}), (\ref{f3}), (\ref{f2}). The transformations 
 (\ref{f7})   
 represent
 the $\beta=\pi/2, 3\pi/2$ rotations of the $O(2,\z)$ group. 
 In the next section we will discuss in detail the action of 
 these duality
 transformations on the space of the solutions of the field 
 equations.
 
 Until now our discussion has been confined to  3D solutions of 
 the form (\ref{e3}). However, it is not difficult to realize that our 
 derivation holds  
 in every situation  where the metric has a diagonal form and the 
 solutions depend only on one spacetime coordinate. In particular 
 our discussion applies to time-dependent  solutions  
 of the form:  
\be \lb{f8}
\ds-dt^2+e^{2\rho(t)}d\x^2+e^{2\n(t)}d\varphi^2,\quad 0\le\varphi
\le 2\pi, \quad 0 \le\x \le 2\pi,\quad \f=\f(t).
\ee
Notice that we consider 3D cosmological solutions with two compact 
spatial
dimensions. 
All the features of the model described in this section rely 
only on the diagonal form of the metric and on its independence from
two of the three spacetime coordinates, so that the generalization to 
a spacetime with metric (\ref{f8}) is straightforward.

Using  the ansatz (\ref{f8}) the field equations (\ref{e2}) become 

\bea 
\ddot{\n}&+&\dot{\n}^2+\dot{\n}\dot{\r}-2\dot{\n}\dot{\f}+
{\la\over k+1}=0,
\nonumber\\
\ddot{\r}&+&\dot{\r}^2+\dot{\n}\dot{\r}-2\dot{\r}\dot{\f}+
{\la\over k+1}=0,
\nonumber\\
\ddot{\n}&+& \ddot{\r}+\dot{\n}^2+\dot{\r}^2-2\ddot{\f}+ 
{4\over k}\dot{\f}^2
+{\la\over k+1} =0,
\nonumber\\
\ddot{\f}&-&\dot{\f}\left(2\dot{\f}-\dot{\n}-\dot{\r}\right)=
\la {k\over 2(k+1)},\nonumber
\eea
where the dots denote the derivative $d/dt$.
Again, these equations are equivalent to the equations of motion 
derived from 
the action
\bc
S=\int dt e^{-2\f+\r+\n}\left(4\dot{\r}\dot{\f}+4\dot{\n}\dot{\f}-
2\dot{\n}
\dot{\r}
+4\left({1\over k}-1\right)\dot{\f}^2+\l^2\right),
\ec
together with the 
Hamiltonian constraint 
\bc 
4\dot{\n}\dot{\f}+4\dot{\f}\dot{\r}-2\dot{\r}\dot{\f}
+4\left({1\over k}-1\right)\dot{\f}^2-\l^2=0.
\ec
Using the redefinitions  (\ref{e7}), the previous equations  can 
be written 
in terms of the new variables $U, Z, Y$ as follows

\bc
S=-\int dt e^{-U}\left[ {k+1\over k+2}\,\dot{U}^2 -{1\over 2}
\left(\dot{Z}^2+\dot{Y}^2\right)-\l^2\right],
\ec
 \bc
 {k+1\over k+2}\,\dot{U}^2 -{1\over 2}
\left(\dot{Z}^2+\dot{Y}^2\right)+\l^2=0.
\ec
The continuous $O(2)$ symmetry (\ref{e9}), (\ref{f1}) and the duality 
symmetries 
 (\ref{f4}), (\ref{f3}), (\ref{f2}) of the field 
equations, discussed previously for the static solution (\ref{e3}),
are also  symmetries of the cosmological solutions  
(\ref{f8}). The duality symmetries of these solutions are now a 
generalization to the generic dilaton gravity theory defined by 
the action 
(\ref{e1}) 
of the scale factor duality \cite{ GV, GAV, TV} of low-energy 
string theory .

\section{Static solutions}
\paragraph{}

In this section we will derive and discuss the most general static, 
circularly symmetric, solution of our 3D dilaton gravity theory. In 
particular, we will also analyze the action of the dualities 
(\ref{f4}), 
(\ref{f3}), (\ref{f2}) on 
the space of the solutions.

The continuous $O(2)$ symmetry is a feature of the field equations 
(\ref{e4}) that is strongly related with their solubility. In fact, 
the field redefinitions (\ref{e7}) not only make the $O(2)$ symmetry 
explicit but also enable us to write 
the  field equations (\ref{e4}) in a form that makes them elementary 
solvable. Varying the action (\ref{e8}) with respect to $Z, Y, U$ and 
using the constraint (\ref{e8a}),
one gets
\bea\lb{g1}
Z''&-&U'Z'=0,\nonumber\\ 
Y''&-&U'Y'=0,\nonumber\\ 
U''&-&U'^2+{k+2\over k+1}\la =0.
\eea 
These equations, supported with the constraint 
(\ref{e8a}),  are equivalent to the field equations (\ref{e4}).
 The system of differential equations (\ref{g1})
 can be  easily integrated. The analytic expressions and the 
 properties of the corresponding solutions  depend on the value of the 
 parameter $k$, therefore we will discuss the various cases 
 separately.
 \subsection{${\bf\{ k<-2\} \cup \{k>-1\} }$}
\paragraph{}
The solutions, expressed in 
terms of the metric variables $\r,\n$ and of the dilaton $\f$, are
\bea \lb{g2}
e^{2\r}&=&A\biggl[\sinh\alpha(\s-\s_0)\biggr]^{2\g+{2\over\kv}}
\biggl[\cosh\alpha(\s-\s_0)\biggr]^{{2\over\kv}-2\g},\nonumber\\
 e^{2\n}&=&B\biggl[\sinh\alpha(\s-\s_0)\biggr]^{2\d+{2\over \kv}}
\biggl[\cosh\alpha(\s-\s_0)\biggr]^{{2\over \kv}-2\d},\nonumber\\
e^{-2\f}&=&C\biggl[\sinh\alpha(\s-\s_0)\biggr]^{-(\g+\d)+{k\over\kv}}
\biggl[\cosh\alpha(\s-\s_0)\biggr]^{\g+\d +{k\over \kv}},
\eea
where
\be\lb{g2a}
 \a=\ha\l \sqrt{\kv\over \ku}
 \ee
  and $A,B,C,\s_0,\g,\d$ are 
integration constants. The constants $A, B, C,\s_0$ are arbitrary.  
$\s_0$ is
irrelevant for our discussion, in the following we will set
$\s_0$=0. The constants $\g$ and $\d$ are not arbitrary, in 
fact the constraint (\ref{e8a}) implies
\be\lb{g3}
\g^2+\d^2+{2\over \ku}\g \d={k\over \kv}.
\ee  
The solutions (\ref{g2}) describe black holes with a regular horizon 
only
for 
\bc
\g={\ku\over\kv},\qquad \d=-{1\over \kv}\, .
\ec
In this case the coordinate transformation
\bc
r=\left({b\over a^2}\right)^{1\over\kv}
\biggl(\cosh\a\s\biggr)^{2\over\kv},\quad
a^2={\la\over (\ku)(\kv)}, 
\ec
with an appropriate choice of the constants $A$, $B$ and $C$,
brings the solutions (\ref{g2}) into the following form:
\bea \lb{g5}
ds^2&=&-\left[(ar)^2-br^{-k}\right]dt^2+\left[(ar)^2-
br^{-k}\right]^{-1}dr^2
+r^2 d\varphi^2, \nonumber\\
\ef&=&(ar)^k, \qquad b\ge 0.
\eea
The constant $b$ appearing in the previous equations is related to the 
Arnowitt-Deser-Misner mass of the black hole by $M=(k+1)a^k b$.
The  solutions (\ref{g5}) are the black hole solutions  first found in 
Ref. \cite{SKL} .
They describe a spacetime with 
anti-de Sitter 
asymptotical behavior. For $k>-1$ and $b\neq 0$ we have 
a curvature singularity, shielded by the event horizon, 
located at $r=0$, whereas the asymptotical region is at $r=\i$. 
For $k<-2$ the curvature singularity is located at 
$r=\i$ whereas the asymptotical region is at $r=0$.  

The duality transformations (\ref{f4}), (\ref{f3}), (\ref{f2}) act on 
the space of the solutions 
relating the black hole solutions (\ref{g5}) to other solutions of the 
field equations, similarly to what happens in 2D dilaton gravity 
theories 
\cite{CM}.
The duality (\ref{f4}) transforms  in  eqs. (\ref{g2})  $\g\ra -\g$,
$\d\ra-\d$. This is equivalent to interchange everywhere in 
eqs. (\ref{g2})
the hyperbolic sine with the hyperbolic cosine.
The solution dual to (\ref{g5}) is therefore characterized by 
$\g=-(\ku)/(\kv), \d=1/ (\kv)$. It can be put in the form (\ref{g5})
(with $b<0$) using the coordinate transformation
\bc
r=\left(-{b\over a^2}\right)^{1\over \kv}
\biggl(\sinh\a\s\biggr)^{2\over \kv}.
\ec
Thus, the duality (\ref{f4}) transforms  a black hole   
(a solution (\ref{g5}) with $b>0$) in a spacetime with a naked 
singularity (a solution (\ref{g5}) with $b<0$).
The duality (\ref{f3}) interchanges $\r$ and $\n$ leaving the dilaton 
invariant ($\g \leftrightarrow \d$ in eqs. (\ref{g2})). The black hole 
solutions (\ref{g5}) are mapped by this duality into the solutions
\bea \lb{g6}
ds^2&=&- r^2 dt^2+\left[(ar)^2-br^{-k}\right]^{-1}dr^2
+\left[(ar)^2-br^{-k}\right] d\varphi^2,\nonumber\\
\ef&=&(ar)^k, \qquad b\ge 0.
\eea
In  the neighborhood
of  $r=r_0= (b/a^2)^{1/(\kv)}$ the line element (\ref{g6}) can be 
brought 
into the form
\be\lb{si}
\ds-dt^2+d\s^2 +h^2\s^2d\varphi^2,
\ee
where $h$  is a constant, which depends only on the parameters 
$a$ and $b$.
$r=r_0$ is just a conical singularity for the metric (\ref{g6}).
With an appropriate choice of the constant $B$ appearing in eqs. 
(\ref{g2}), 
one can have 
$h=1$, removing the conical singularity,  so that the point 
$r=r_0$ can
be considered 
as the origin of the radial coordinate. In this case  the 
solution (\ref{g6}) with $k>-1$  
describes  a regular spacetime, which is 
asymptotically anti-de Sitter. It is important to stress  that the 
spacetime with the conical singularity {\sl not}   this 
regular solution is the dual of the black hole (\ref{g5}). 

The duality (\ref{f2}) transforms in eqs. (\ref{g2})
$\g\ra-\d, \d\ra-\g$. This transformation 
 can be obtained from the product of the dualities 
 (\ref{f4}) and (\ref{f3}). It is easy to see  that it 
maps 
the  black hole solutions (\ref{g5}) into the solutions (\ref{g6}) 
with $b<0$.
In this case the solutions describe a spacetime with a naked 
singularity.

\subsection{$\bf{-2<k<-1}$}
\paragraph{}
The equations (\ref{g1}) are now solved by the expressions 
(\ref{g2}) with 
the hyperbolic functions replaced by the corresponding trigonometric 
ones and with 
$\a={\l\over 2}\sqrt{|{\kv\over \ku}|}$. The integration constants 
$\g$ and $\d$ have 
to satisfy the same constraint (\ref{g3}).
The solutions characterized by 
\bc
\g={\ku\over\kv},\qquad \d=-{1\over \kv},
\ec
now describe  a spacetime with  a cosmological horizon.
 In fact performing the coordinate transformation 
\bc
r=\left({b\over a^2}\right)^{1\over\kv}
\biggl(\cos\a\s\biggr)^{2\over \kv},
\ec
these solutions become 
\bea \lb{g7}
ds^2&=&-\left[br^{-k}-(ar)^2\right]dt^2+\left[br^{-k}-
(ar)^2\right]^{-1}dr^2
+r^2 d\varphi^2,\nonumber\\
\ef&=&(ar)^k, \qquad b\ge 0 .
\eea
The dualities (\ref{f4}), (\ref{f3}), (\ref{f2}) act on the solution
(\ref{g7}) in a way  similar to that described in sect. 3.1. 
One must simply replace 
everywhere hyperbolic  with trigonometric functions. In particular, 
the 
duality (\ref{f4}) transforms in (\ref{g7}) $b\ra-b$, i.e the dual of 
(\ref{g7}) is a singular spacetime with no horizons. 
The duality (\ref{f3}) 
transforms the spacetime (\ref{g7}) into  
\bea
ds^2&=&- r^2 dt^2+\left[br^{-k}-(ar)^2\right]^{-1}dr^2
+\left[br^{-k}-(ar)^2\right] d\varphi^2,\nonumber\\
\ef&=&(ar)^k,\nonumber
\eea
with $b\ge 0$
whereas (\ref{f2}) transforms (\ref{g7}) into the same solution 
but with $b\le0$.
 
\subsection{${\bf k=\i}$}
\paragraph{}
This case corresponds to 3D  string effective theory. The static, 
circularly  symmetric, solutions are obtained by taking
in eqs. (\ref{g2}), (\ref{g2a}) and  (\ref{g3}) (we set there 
$A=1,B=\l^{-2}$) the $k\ra \i$ limit:
\bea \lb{g8}
e^{2\r}&=&\left(\tanh{\l\over2}\s\right)^{2\g},
\nonumber\\
e^{2\n}&=&\l^{-2}\left(\tanh{\l\over2}\s\right)^{2\d},
\nonumber\\
e^{-2\f}&=&C\left(\sinh{\l\over2}\s\right)^{1-(\g+\d)}
\left(\cosh{\l\over2}\s\right)^{\g+\d +1},
\eea
with $\d^2+\g^2=1$.
The change of variable 
\bc
{r\over M}=\left(\cosh {\l\over 2} \s\right)^2,
\ec
enables us to write (\ref{g8}) as follows
\bea\lb{g9}
ds^2&=&-\left(1-{M\over r}\right)^\g dt^2+\left(1-
{M\over r}\right)^{-1}
 {1\over\la r^2} \, dr^2 +
\l^{-2}\left(1-{M\over r}\right)^\d d\varphi^2, \nonumber\\
\ef&=&\l r\left(1-{M\over r}\right)^{(1-\g-\d)/2},\qquad \g^2+\d^2=1 .
\eea
Differently from the general case, this solution  describes a
spacetime that is asymptotically flat with topology $R^{1,1}
\times S^1$ 
(we consider here a compact coordinate $\varphi$, our consideration 
can be 
easily generalized to the case of a noncompact $\varphi$). 
This implies
that string effective theory in three dimensions does not admit black 
hole  but only black string solutions.
This result is 
already known \cite{HH,GH} and eq. (\ref{g9}), being 
the general static, 
circularly symmetric, solution of 3D string effective theory, gives a 
straightforward prove of it.
 The black string of Ref. \cite{HH} is 
obtained from eq. (\ref{g9}) (or equivalently from (\ref{g8})) 
by setting 
there $\g=1,\d=0$:
\bea\lb{h1}
ds&=&-\left(1-{M\over r}\right) dt^2+\left(1-{M\over r}\right)^{-1}
{1\over\la r^2} dr^2 +\l^{-2} d\varphi^2,\nonumber\\
\ef&=&\l r, \qquad M\ge 0.
\eea

For generic values of the parameters $\g$ and $\d$ the solution 
(\ref{g9})
 presents  a curvature singularity at $r=M$. For $\g=-1, \d=0$ the 
 solution has the form (\ref{h1}) but with $M\le0$. In this case the 
 spacetime has  a curvature singularity at $r=0$ not shielded by an 
event horizon. For $\g=0$ and $\d=\pm1$ the solutions can be put 
into the following 
form
\bea \lb{h1a}
ds^2&=&-\l^{-2}dt^2+\left(1-{M\over r}\right)^{-1}
{1\over\la r^2}\,dr^2 + \left(1-{M\over r}\right)  d\varphi^2,,
\nonumber\\
\ef&=&\l r,
\eea
where $M\ge 0$ for $\d=1$ and  $M\le 0$ for $\d=-1$.
For $M>0$, $r=M$ is a conical singularity of the metric. In fact,
in the neighborhood of  $r=M$ the line element (\ref{h1a}) reduces 
to the form
(\ref{si}).
 We can remove this conical 
singularity by choosing appropriately either the integration 
constants or 
the period of the compact coordinate $\varphi$.
Near $r=M$ the spacetime   has  the local topology of $R^{1,2}$. 
Thus, $r=M$ 
can be considered as 
the origin of a polar coordinate system of  the  plane. In this case 
the metric 
(\ref{h1a}) 
describes therefore  a 3D  regular spacetime with local topology 
$R^{1,1} \times S^1$ for $r\ra \i$  and $ R^{1,2}$ for $r\ra M$.
Conversely, for $M<0$ the line element (\ref{h1a}) has a curvature 
singularity at $r=0$.   

As for the general case, using the duality transformations 
(\ref{f6})
one can relate the black string solution (\ref{h1}) to the  various  
spacetimes discussed in this section. In particular, the first duality 
in eqs. (\ref {f6}), which interchanges the 
hyperbolic sine and cosine, is the well-known target space 
duality of 2D 
string theory that interchanges the horizon and the singularity 
of the 2D 
black hole geometry \cite{GPR,AG,VV}. This transformation  maps in 
eqs. (\ref{g9}) the 
solutions with positive $M$ in 
those with $M$ negative. The second duality in eqs. (\ref{f6}) 
maps the black string  into the  spacetime with a conical singularity 
described by the  line element (\ref{h1a}) with $M>0$.
 Finally, the third duality in eqs. (\ref{f6}) relates the black 
 string 
solution to 
the line element (\ref{h1a}) with $M <0$, which describes a spacetime 
with a curvature singularity.  

\subsection {${\bf k=0,-2}$}
\paragraph{}
For  $k=0$, which corresponds to 3D general relativity, the dilaton 
must 
be constant  and eq. 
(\ref{g3}) 
implies $\d=-\g$. Introducing the new 
radial coordinate $r={\sqrt{b}\over a}\,\cosh\a\s,$ 
the general solution takes the form:
\bc
ds^2=- (ar)^{1-2\g}\left[(ar)^2-b\right]^{\ha+\g}dt^2+\left[(ar)^2-
b\right]^{-1}dr^2
+ (ar)^{1+2\g}\left[(ar)^2-b\right]^{\ha-\g}a^{-2} d\varphi^2, 
\ec
where $\g$ now is an arbitrary constant.
We have regular black holes solutions only for 
$\g=1/2$, which gives the BTZ black hole \cite{BTZ}  
\bc 
\ds-\left[(ar)^2-b\right]dt^2+\left[(ar)^2-b\right]^{-1}dr^2
+r^2 d\varphi^2. \nonumber
\ec
Here the only relevant duality is  (\ref{f3}) that interchanges  the 
BTZ black hole solutions with $b>0$ with those with $b<0$.

For $k=-2$  the solutions to the field equations (\ref{g1}) are 
\bea
e^{2\r}&=&A\exp\left({- \la \s^2\over 2}\right)(\l\s)^{\d-\g},
\nonumber\\
e^{2\n}&=&B\exp\left({- \la \s^2\over 2}\right)(\l\s)^{\d+\g,}
\nonumber\\
e^{2\r}&=&C\exp\left({ \la \s^2\over 2}\right)(\l\s)^{1-\d},
\nonumber\\
\g^2&+&2\d-3=0.\nonumber
\eea
For generic values of the parameters $\g$ and $\d$ the corresponding 
metric has  curvature singularities both at $\s=0$ 
and $\s=\i$. For $\d=1$, $\g=\pm 1$, $\s=0$ becomes a coordinate 
singularity. The duality transformations (\ref{f4}) and (\ref{f2})
 become singular for $k=2$. The only transformation that  is a  
 symmetry of the field equations 
 is now given by eq. (\ref{f3}). This transformation  
flips the sign of $\g$ in the 
solutions.

\section{Cosmological solutions}
\paragraph{}
Let us now consider time-dependent solutions that have the 
form (\ref{f8}).
The field equations and the hamiltonian constraint can be written as 
follows 

\bea\lb{l1}
\ddot{Z}&-&\dot{U}\dot{Z}=0,\nonumber\\ 
\ddot{Y}&-&\dot{Y}\dot{Z}=0,\nonumber\\ 
\ddot{U}&-&\dot{U}^2-{k+2\over k+1}\la=0,\nonumber\\
\l^2&+&{k+1\over k+2}\,\dot{U}^2 -{1\over 2}
\left(\dot{Z}^2+\dot{Y}^2\right)=0 .
\eea 
This system of differential equations can be easily solved. 
In fact, it 
is related to the static case (eqs.(\ref{g1})) by the substitution
 $\s\ra it$.
In the same way as for the static solutions, we will consider 
separately
 the models 
with different values of the parameter $k$ . 

\subsection{${\bf\{ k<-2\} \cup \{ k>-1\} }$}
\paragraph{}
Expressed in terms of the metric functions $\r$,$\n$ and of the 
dilaton $\f$, the general 
solution of the eqs. (\ref{l1}) is
\bea \lb{l2}
e^{2\r}&=&A\biggl[\sin\alpha(t-t_0)\biggr]^{2\g+{2\over \kv}}
\biggl[\cos\alpha(t-t_0)\biggr]^{{2\over \kv}-2\g},\nonumber\\
 e^{2\n}&=&B\biggl[\sin\alpha(t-t_0)\biggr]^{2\d+{2\over \kv}}
\biggl[\cos\alpha(t-t_0)\biggr]^{{2\over\kv}-2\d},\nonumber\\
e^{-2\f}&=&C\biggl[\sin\alpha(t-t_0)\biggr]^{-(\g+\d)+{k\over\kv}}
\biggl[\cos\alpha(t-t_0)\biggr]^{\g+\d +{k\over \kv}}.
\eea
In these equations $\a$ is given by eq. (\ref{g2a}) and the 
constants $\g$ 
and $\d$ have 
to satisfy eq. (\ref{g3}).  
 $A,B,C,t_0$ are arbitrary constant, which in the following we will  
 be 
 fixed by setting
$A=B=\a^{-2}, C=1$ and $t_0$=0. 

The solutions (\ref{l2}) have the typical form of the string 
cosmological 
solutions that have been already discussed in the literature 
\cite{TV,MM,CC}. The surfaces 
with $t=const$ in the line element (\ref{f8}) have topology $S^1\times 
S^1$ and the scale factors 
$\exp(2\r)$ and $\exp(2\n)$ have to be considered as the 
(time-dependent) 
radii of the two 
one-spheres $S^1$. In general, the scale factors are periodic, 
with period  
$2\pi/\a$. For generic values of $\g$ and $\d$ the line element  
has curvature  singularities  both at $t=n\pi/\a$ and 
$t=(n+1)\pi/2\a$, ($n=0,1,2 ..$).
For $\g=(\ku)/(\kv), \d=-1/(\kv)$ the solutions (\ref{l2}) become
\bea \lb{l3}
e^{2\r}&=&\a^{-2}\left(\sin\alpha t\right)^2
\left(\cos\alpha t\right)^{-2k/(\kv)},\nonumber\\
 e^{2\n}&=&\a^{-2}\left(\cos\alpha t\right)^{4/(\kv)},\nonumber\\
e^{-2\f}&=&\left(\cos\alpha t\right)^{2k/(\kv)}.
\eea
Though $t=(n+1)\pi/2\a$ is still a curvature singularity,
$t=n\pi/\a$ becomes  a coordinate singularity of the metric. 
In fact, in the 
neighborhood of this point the line element ({\ref{f8}}) 
reduces to the form
\bc
\ds-dt^2 +t^2 d\chi^2 +\a^{-2}d\varphi^2.
\ec
The topology of the spacetime is locally $R^{1,1} \times S^1$ and the 
surface $t=0$ is homotopic to $S^1$ and a point. 
For $\g=-1/(\kv), \d=(\ku)/(\kv)$ we have the same situation but 
the scale factors of the two one-spheres are interchanged. 
The solution has 
now the following form:
\bea \lb{l4}
e^{2\r}&=&\a^{-2}\left(\cos\alpha t\right)^{4/(\kv)},\nonumber\\
 e^{2\n}&=& \a^{-2}\left(\sin\alpha t\right)^2
\left(\cos\alpha t\right)^{-2k/(\kv)},\nonumber\\
e^{-2\f}&=&\left(\cos\alpha t\right)^{2k/(\kv)}.
\eea  

The action of the duality transformations (\ref{f4}), (\ref{f3}),
(\ref{f2}) on 
the space of the solutions is similar to that described in sec. 3.1 
for the 
static solutions. In particular, the duality (\ref{f4}) is the 
3D analogue of the 2D scale factor duality discussed in Ref. 
\cite{CC}.
It interchanges   
everywhere in eqs. (\ref{l2}) the sine with the cosine, so that 
for generic $\g$
and $\d$  it interchanges one with the other the two curvature 
singularities at
$t=(n+1)\pi/2\a$ and $t=n\pi/\a$. In  the case of solutions 
(\ref{l3}), (\ref{l4}) 
it  interchanges the curvature singularity at $t=(n+1)\pi/2\a$ 
and the 
coordinate singularity at $t=n\pi/\a$. The duality (\ref{f3}) 
interchanges the 
scale factor $\exp 2\r$  with the scale factor $\exp2\n$ 
leaving the dilaton 
invariant.
Solution (\ref{l3}) is mapped by this transformation into solution 
(\ref{l4}) 
and {\sl vice versa}.
Finally, the duality (\ref{f2}), being the product of dualities 
(\ref{f4}), 
(\ref{f3}) first interchanges the sine and the cosine then 
interchanges the scale 
factors of the two $S^1$.

\subsection{${\bf -2< k<-1}$}
\paragraph{}
The solutions to the field equations (\ref{l1}) can now be given 
in terms 
of hyperbolic functions of argument $\a t$, with 
$\a=(\l/2)\sqrt{|(\kv)/(\ku)|}$. They have the form (\ref{l2}) 
with the 
trigonometric functions replaced by the corresponding 
hyperbolic ones. The main 
difference between the present case and that of Sect. (4.1) is that 
now the 
cosmological solutions are not any more periodic. 
The generic solution has only a curvature singularity at $t=0$.
For the special values of the parameters $\g,\d$ leading to the 
solutions (\ref{l3}), (\ref{l4}), the curvature singularity at $t=0$ 
becomes a coordinate singularity. In this case the solutions describe
a regular spacetime. 
 All the main features of the solutions described in the previous 
subsection, 
concerning the action of 
the duality 
transformations on the space of the solutions, can be easily 
translated 
to the present case.

\subsection{${\bf k=\i}$}
\paragraph{}
Performing the limit $k\ra \i$ in eqs. (\ref{l2}) one obtains the 
cosmological 
solutions of 3D string effective theory:
 \bea \lb{l5}
e^{2\r}&=&\a^{-2}\left(\tan\alpha t\right)^{2\g},
\nonumber\\
e^{2\n}&=&\a^{-2}\left(\tan\alpha t\right)^{2\d},
\nonumber\\
e^{-2\f}&=&\left(\sin\alpha t\right)^{-(\g+\d)+1}
\left(\cos\alpha t\right)^{\g+\d +1},\nonumber
\eea
with $\d^2+\g^2=1$.
The metric has the same singularities as in the general case. 
For generic values of the parameters $\g$ and $\d$  we have two 
curvature 
singularities whereas for $\g=1,\d=0$ or $\g=0, \d=1$ the 
singularity at 
$t=n\pi/\a$ becomes a coordinate singularity. For these values of the 
parameters $\g$ and $\d$ the radius of one $S^1$ is constant and the 
spacetime has topology $H\times S^1$ where $H$ describes the 
well-known 
cosmological solution of 2D string effective theory \cite{MM,TV}.  
Moreover, the solution with $\g=1,\d=0$ is related to that with 
$\g=0,\d=1$ by the second duality transformation in eqs. (\ref{f6}), 
which interchanges the 
scale factors of the two $S^1$. The first duality in eqs. (\ref{f6}) 
is the 3D 
analogue of the scale factor duality of  2D string effective 
theory, to 
which it reduces when either $\r$ or $\n$ is constant. Finally, 
the third duality 
in eqs. (\ref{f6}) is the product of the first two. 
 
\subsection{${\bf k=0,-2}$}
\paragraph{}
For $k=0$ the dilaton must be constant and the equations (\ref{l2}) 
yield
\bea
e^{2\r}&=&{4\over \la}\left(\sin{\l\over 2}t\right)^{2\g+1}\left(\cos 
{\l\over 
2}t\right)^{1-2\g},\nonumber\\
e^{2\n}&=&{4\over \la}\left(\sin{\l\over 2}t\right)^{1-2\g}\left(\cos 
{\l\over 
2}t\right)^{1+2\g},\nonumber
\eea
where $\g$ now is an arbitrary integration constant.
The previous equations give the cosmological solutions of 3D general 
relativity with a cosmological constant. For generic $\g$ we have, 
again, two curvature singularities. 
For $\g=\pm 1/2$ one of these becomes a coordinate singularity. The 
dualities (\ref{f4}), (\ref{f3}), (\ref{f2}) now reduce to one single 
transformation, which transforms $\g\ra-\g$ or equivalently 
interchanges 
the sine  and the cosine in the solutions.

For $k=-2$ the solutions to the system (\ref{l1}) can be written as 
follows
\bea
e^{2\r}&=&\l^{-2}\exp\left({\la t^2\over 2}\right)(\l t)^{\d-\g},
\nonumber\\
e^{2\n}&=&\l^{-2}\exp\left({\la t^2\over 2}\right)(\l t)^{\d+\g},
\nonumber\\
\ef&=& \exp\left({-\la t^2\over 2}\right) (\l t)^{1-\d},\nonumber
\eea
\bc
\g^2+2\d-3=0.
\ec
For generic $\g ,\d$ these solutions have a curvature singularity both 
at $t=0$ and $t=\i$. For $\d=1, \g=\pm1$, $t=0$ becomes a coordinate 
singularity.   The only transformation that  is a duality of 
the field equations is now given by eq. (\ref{f3}), which flips 
the sign of $\g$ 
in the 
solutions.

\section{ Summary and outlook}
\paragraph{}
In this paper we have discussed in detail the duality symmetries of 
3D dilaton gravity theories. We have found that for  
spacetime metrics with a diagonal form the independence of  
the solutions 
from two spacetime coordinates implies the existence of a 
continuous $O(2)$ and therefore a discrete $O(2,\z)$,
symmetry of the reduced field equations. 
In the case of 3D string effective theory ($k=\i$)
this represents a full field theoretical derivation of the 
discrete $O(2,\z)$ duality symmetry, appearing as subgroup of the 
target space duality group of 3D string theory. Our derivation, 
being based on purely field theoretical arguments, holds for 3D 
Brans-Dicke 
dilaton gravity theories but it is not clear
if it can be generalized to these theories in generic $d$ dimensions.
 The results of ref. \cite{JL}  indicate that this is 
possible at least for  homogeneous and isotropic  cosmological 
solutions. 

From a more conventional point of view, the existence of 
duality symmetries made it possible to unfold fully the richness
of the spectrum of the solution of 3D dilaton gravity theories.
Until now, only few of these solutions were known,
mainly the black hole solutions of the generic theory, the black
string solution of the string effective theory and other 
solutions of dilaton gravity coupled to antisymmetric tensor fields.
We have been able to write down  general   static and 
cosmological solutions of 3D dilaton gravity.
Surprisingly enough, we have found that 3D string effective theory
has static solutions different from the black string. These solutions
describe either a regular spacetime or a spacetime with a 
conical singularity. Besides, the solutions with the conical
singularity are dual to the black string solutions.
Presently, we  do not know if this is just a result of
string effective theory or if it has a counterpart in terms 
of exact string backgrounds, i.e if these 3D spacetimes with conical 
singularities have an interpretation as conformal field theories.
Exact 3D string backgrounds with similar conical singularities have 
been found by 
Horne and Horowitz in the context of Wess-Zumino-Witten models 
\cite{HH}. These  
 are also solutions of 3D string effective theory with an
antisymmetric tensor. We do not know if 
these  solutions with axion charge are related to those we have found 
in this paper. 
In order to answer to this question, one should be able to find the 
general static solution of 3D dilaton gravity coupled to the 
antisymmetric 
tensor.
We leave this issue to further investigations.                        

\end{document}